\documentclass[preprint2]{aastex}

\shorttitle{Globular Cluster Populations of HCGs}
\shortauthors{Barkhouse, West, \& Bothun}

\begin{document}

\title{Globular Cluster Population of Hickson Compact
Group 22a and 90c\altaffilmark{1}}

\author{Wayne A. Barkhouse} 
\affil{Department of Astronomy \& Astrophysics, University of Toronto, 
Toronto, Ontario, Canada, M5S 3H8} 
\email{barkhous@astro.utoronto.ca}

\author{Michael J. West}
\affil{Department of Physics \& Astronomy, University of Hawaii at Hilo, 
Hilo, HI 96720}
\email{west@bohr.uhh.hawaii.edu}

\and
\author{Gregory D. Bothun}
\affil{University of Oregon, Department of Physics, Eugene, OR,
USA, 97403}
\email{nuts@moo.uoregon.edu}

\altaffiltext{1}{Based on observations collected at the European Southern 
Observatory, Chile.}

\begin{abstract}

We present the first measurement of     
the globular cluster populations of galaxies in Hickson compact 
groups, in order to investigate the effect of 
these high density environments on the formation and 
evolution of globular cluster systems.
Based on V and R band images that we obtained 
of HCG 22a and HCG 90c with the ESO New Technology Telescope (NTT), 
we find a total globular
cluster population of $1590\pm 854$ for HCG 22a and $2136\pm 718$ for 90c. 
The specific frequency for HCG 22a was found to be $S_{N}=1.9\pm 1.0$ and 
$S_{N}= 3.4\pm 1.1$ for HCG 90c. A power-law fit to the globular cluster 
radial profile of HCG 22a yields $\sigma\sim R^{-2.01\pm 0.30}$ and for 
HCG 90c we found $\sigma \sim R^{-1.20\pm0.16}$. A comparison of the globular
cluster radial profiles with the surface brightness of the parent galaxy 
shows that the globular cluster systems are at least as extended as the 
halo light. The measured values for the specific frequency are consistent 
with a scenario in which the host galaxies were in a low density 
``field-like'' environment when they formed their globular cluster systems.
 
\end{abstract}


\section{Introduction}\label{sec:intro}
Tremendous progress has been made in recent years in the detection of 
globular cluster (GC) populations around galaxies beyond the Local Group
\citep[e.g.,][]{har91,bla99}. The globular cluster populations of well 
over 100 galaxies up to distances $\gtrsim 100h^{-1}$ Mpc have now 
been studied in various environments in the hope of gaining further insight 
into galaxy formation and evolution. Studies have shown that the number 
of globular clusters associated with any particular galaxy varies widely, 
from none for M32 to $\sim 20,000$ for the cD galaxy M87 
\citep[e.g.,][]{har91}.

A useful parameter for quantifying the total number of globular clusters per
galaxy is the specific frequency, $S_{N}=N_{tot}10^{0.4(M_{V}+15)}$,
which is defined as the total number of globular clusters, $N_{tot}$, 
per unit galaxy luminosity \citep[normalized to $M_{V}=-15$;][]{har81}. 
Previous studies have shown that globular cluster populations vary 
systematically with galaxy type and environment \citep{har91,west94}. 
Characteristic values of $S_{N}$ range from $S_{N}\le 1$ for field spiral 
galaxies, to $S_{N}\simeq 2-4$ for field ellipticals, to $S_{N}\simeq 5-7$ 
for ellipticals in dense environments, and up to $S_{N}\simeq 10-20$ for 
supergiant ellipticals at the centers of rich clusters. (We note that 
normalization with respect to bulge luminosity for spiral galaxies would 
increase the value of $S_{N}$, see C\^{o}te et al. 2000.) 
Yet, which factor has the greatest influence on globular cluster
formation---galaxy type or environment---is unclear. 

Several competing theories have been proposed to explain the origin of 
the observed variations in globular cluster systems of different galaxies. 
Among these are: galaxy mergers \citep{sch86,ash92}, accretion 
via tidal stripping or merging (C\^{o}te, Marzke, \& West 1998; C\^{o}te 
et al. 2000), 
and two-stage collapse models (Forbes, Brodie, \& Grillmair 1997).
At present, it is unclear whether a single mechanism, or a combination of 
proposed formation scenarios, can explain the properties of globular 
cluster systems \citep{kis00}. Determining which scenario most accurately 
describes the formation of globular clusters will require the investigation 
of GC systems in a wide range of environments. 

In this paper we present observations of the globular cluster population of 
two galaxies in Hickson compact groups; HCG 22a (NGC 1199) and HCG 90c 
(NGC 7173). Compact groups are interesting because, although they are small 
(typically 3--7 member galaxies per group), they are extremely dense systems,
comparable in density to the cores of rich Abell clusters 
\citep[e.g.,][]{hic82,hic92}. This makes them a useful tool for probing the
influence of environment on globular cluster formation.

Figure~\ref{22a_image} and \ref{90c_image} show V band images of HCG 22a 
and 90c, respectively. Both HCG 22 and 90 have been detected by ROSAT 
and each group has been shown to be a weak X-ray source \citep{pon96}. 
Table~\ref{gal_data} summarizes the general properties of these systems. 
We discuss each in detail below.

\subsection{Hickson Compact Group 22}

HCG 22 initially was thought to contain five members (three ellipticals 
and two spirals) but later redshift measurements determined that two of 
the elliptical galaxies (HCG 22d and 22e) were most likely background 
galaxies since their heliocentric radial velocities are 
$\sim 7000~\mbox{km}~\mbox{s}^{-1}$ greater than the remaining group 
members \citep{hic92}. The median heliocentric radial velocity of the 
three bona fide group members is $2700~\mbox{km}~\mbox{s}^{-1}$, with 
a velocity dispersion of $43.7~\mbox{km}~\mbox{s}^{-1}$.

From Table~\ref{gal_data}, the heliocentric velocity of HCG 22a is
$2705~\mbox{km}~\mbox{s}^{-1}$ which indicates that this galaxy is most
likely located near the group center of mass. \citet{hic92} estimated 
the mass-to-light ratio of HCG 22 to be $M/L\sim 1.3h^{-1} M_{\odot}/L_{\odot}$ and a projected median galaxy-to-galaxy separation of 
$26.9h^{-1}\mbox{kpc}$. The group crossing time, defined as the ratio of 
the crossing time to the age of the universe, is $Ht_{c}= 0.1905$
which is greater than the median value of $Ht_{c}= 0.016$ determined for 
all of the Hickson compact groups combined \citep{hic92}. This suggests 
that galaxy mergers and interactions may be less important for HCG 22 than 
for other compact groups with shorter crossing times.

\subsection{Hickson Compact Group 90}

HCG 90 consists of four bright galaxies (two ellipticals, a spiral, and an
irregular) with a median heliocentric group radial velocity of
$\sim 2640~\mbox{km}~\mbox{s}^{-1}$ and a velocity dispersion 
of $\sim 100~\mbox{km}~\mbox{s}^{-1}$ \citep{hic92}. The mass-to-light 
ratio of HCG 90 is given by \citet{hic92} as 
$M/L\sim 12.3h^{-1} M_{\odot}/L_{\odot}$ and a median projected galaxy 
separation of $29.5h^{-1}\mbox{kpc}$. Table~\ref{gal_data} lists the 
heliocentric radial velocity of HCG 90c as $2696~\mbox{km}~\mbox{s}^{-1}$, 
which indicates that this galaxy is probably close to the center
of mass of the group. 

The group crossing time was estimated by \citet{hic92} to be $Ht_{c}= 0.0224$
, which is only $\sim 12\%$ of the estimated value for HCG 22. This 
suggests that galaxy mergers and interactions are likely to have been 
more important in HCG 90 than for HCG 22 and it is interesting to note 
that HCG 90 contains a galaxy (HCG 90d) that is clearly interacting with 
other group members \citep{lon94,pla99}.  

In this study, we have measured the globular cluster specific frequency of
HCG 22a and HCG 90c. In \S~\ref{sec:obs} we present the observations and 
data reductions. 
Section~\ref{sec:results} presents the results for HCG 22a and 90c in 
terms of their globular cluster luminosity functions, radial distributions, 
and specific frequencies. In \S~\ref{sec:dis} we discuss the results of 
this study in the context of globular cluster formation and the nature 
of Hickson compact groups.
  
Throughout this paper, unless otherwise specified, we use 
$H_{0}= 50~\mbox{km}~\mbox{s}^{-1}\mbox{Mpc}^{-1}$. 

\section{Observations and Data Reductions}\label{sec:obs}

Observations of HCG 22a and 90c were obtained on the nights of 
October 9 and 10, 1993 using the 3.5m New Technology Telescope (NTT) 
operated by the European Southern Observatory (ESO) in La Silla,
Chile. All images were taken with the ESO Multi-Mode Instrument (EMMI) 
with a Loral 2048 CCD mounted on the red arm. The Loral 2048 CCD chip 
consists of 2048$\times$2048 15$\micron$ pixels, giving a scale 
of $0.29\arcsec$ per pixel at the f/11 Nasmyth focus. Technical limitations 
during the observing run limited the effective area of the chip to 
1700$\times$1700 pixels, resulting in a projected sky coverage of 
$8.2\arcmin\times8.2\arcmin$.
 
The observations consisted of $5\times 900$ second exposures in the Cousins 
V and R passband for HCG 22a, and $6\times 900$ seconds in V and R for
HCG 90c. Since globular clusters around these galaxies can be detected 
only as a statistical excess of star-like objects, a series of 
$5\times 900$ second exposures in V and R were also taken of a control 
field located within $3\arcdeg$ of the target galaxies in order to 
ascertain the expected number of contaminating Galactic stars and 
unresolved faint galaxies along the line of sight. The seeing varied 
during the observing run from $0.87\arcsec$ to $0.98\arcsec$ (FWHM), 
as measured from the individual galaxy frames.

Processing of the raw images was conducted within the 
IRAF\footnote{IRAF is distributed by National Optical Astronomy 
Observatories, which are operated by the Association of Universities 
for Research in Astronomy, Inc., under cooperative agreement with the 
National Science Foundation.} environment and involved the standard 
procedure of bias correction and flat fielding using dome flats. After 
this initial processing, each set of individual galaxy and control 
field images were registered to their common coordinate system and 
median combined to yield a higher S/N image. Instrumental magnitudes 
were transformed to the standard UBVRI system by measuring several 
standard stars from \citet{lan92}.

At the distance of the target galaxies ($\sim 54$~Mpc), and with the 
typical seeing of the combined frames ($\sim 0.9\arcsec$), individual 
globular clusters are unresolved and appear as an apparent excess of 
star-like objects around their parent galaxy. In order to maximize the 
detection of individual globular clusters and to perform accurate 
photometry, the light from each host galaxy was removed. This was done 
by using the STSDAS package in IRAF to model the light distribution of 
each galaxy using the tasks {\tt ellipse} and {\tt bmodel}, which 
fit isophotes and then generate models of each galaxy's surface 
brightness profile. The constructed galaxy models were then subtracted 
from the appropriate image.
                         
The galaxy-subtracted image of HCG 22a revealed a large, central, diffuse
structure which is the residual of a known central dust lane 
(Franx, Illingworth, \& Heckman 1989). In order to remove this feature, 
the galaxy-subtracted image was median filtered using a ring median 
filter \citep{sec95} having an inner radius of 6.36 pixels and a width 
of 1 pixel. The resultant smoothed image was subtracted from the previous 
image, which helped to remove most of the dust lane feature. A second 
rectangular median filter of dimension 10$\times$10 pixels was applied 
to the resultant image and this smoothed image was subtracted from the 
previous image. The galaxy light 
removal procedure was applied equally to the V and R images. 

The HCG 90c galaxy model was subtracted from the parent image and the 
result was median filtered using a ring filter with the same dimension 
as that used for HCG 22a. The resultant smoothed image was then subtracted 
from the previous image, adequately removing all galaxy light. This 
procedure was applied equally to the V and R images.

Object detection and classification was conducted using the Faint 
Object Classification and Analysis System (FOCAS) within the IRAF 
environment \citep{jar81,val93}. A threshold of $2.5\sigma$ above 
background and a minimum object area of six pixels was used for the 
detection criteria. The detected objects were classified by FOCAS as 
star-like or galaxy-like by comparing a model PSF, generated by the 
average profile of a minimum of ten isolated, unsaturated stars, with 
the light profile of each object. The object list was then culled of 
saturated and galaxy-like objects. A master catalog of stellar-like 
objects detected in both the V and R images was generated for each galaxy 
and the control field. Only objects that were matched in both filters 
were included in the final sample, thereby greatly reducing the possibility 
of false detections of objects. DAOPHOT was used to perform the photometry 
on all detected objects.

Extinction corrections (Burstein \& Heiles 1982) were determined 
to be negligible for HCG 90 and the control field ($A_{B}=0.0$). 
For HCG 22, we determined $A_{V}=0.08$ magnitude and $A_{R}=0.06$ magnitude. 

\subsection{Completeness Function}
     
Since the globular cluster population of HCG 22a and 90c are 
detected only at very faint magnitudes ($V\ge 22$), the probability 
of detecting an object as a function of magnitude must be determined 
as accurately as possible. A ``completeness'' function was calculated 
by adding artificial stars (generated using the DAOPHOT routine 
{\tt Addstar}) to the galaxy and control field images and then re-running 
the FOCAS routines. The fraction of detected artificial stars as a function 
of magnitude was then used to derive the completeness function.    

Artificial stars were added to both the V and R images using the same 
spatial coordinates and with $\vr=0.4$ (the approximate color of Milky 
Way globular clusters; see Peterson 1993). FOCAS was then run {\em exactly} 
in the same manner as that used for real objects, including keeping only 
those objects detected in both the V and R images. 
  
The photometric magnitude limit of the data is defined as the magnitude 
at which the probability for detecting an object equals 50\% \citep{har90}
. For magnitudes fainter than this limit, object information is much 
less reliable and thus all such objects were discarded from further 
analysis. The completeness limit for HCG 22a was determined to be 
$V=24.9$.
 
Since noise and degree of crowding increases towards the center of 
the subtracted galaxy, the completeness function will vary with radial 
distance from the galaxy center. Artificial stars were placed at different 
radii from the galaxy center and the completeness limit, as a function of 
radial distance, was measured. The completeness function was measured 
as a function of both magnitude and position from the center of the subtracted 
galaxy. 

For HCG 22a, it was determined that for 
radii $<$ 70 pixels, the completeness limit was several magnitudes brighter 
than for radii $> 70$ pixels (part of this is due to the small disk-like 
artifact found at the center of HCG 22a). All objects found within 70 pixels 
of the galaxy center were discarded and this inner region was not used in 
any further analysis. For the area outside of this region, the 
completeness function was found to be approximately the same.
 
For HCG 90c, the area within 60 pixels of the galaxy center was 
determined to be unusable due to its bright magnitude completeness limit. 
This region was masked out and not used in any further analysis. 
The completeness limit of the area outside of this region was determined 
to be $V=24.8$.

For the control field, artificial star experiments indicate a completeness
limit of $V=24.6$.

Photometric errors were calculated by determining the standard deviation 
of the magnitude difference between added and recovered artificial stars. 
The root-mean-square of the random error for HCG 22a ranged from 
$\sigma_{rms}\approx 0.01$ at $V=22$ to $\sigma_{rms}\approx 0.06$ 
at $V=24.6$. For HCG 90c, $\sigma_{rms}\approx 0.01$ at $V=22$ and increased 
to $\sigma_{rms}\approx 0.09$ at $V=24.6$.

\subsection{Selection Criteria}

The detection of globular cluster candidates can be improved using 
criteria such as magnitude and shape parameters which are characteristic 
of globular clusters in general. Since objects detected in both galaxy 
fields must be compared to objects detected in the control field, all 
objects in the galaxy image catalogs with magnitudes fainter than $V=24.6$ 
(the completeness limit of the control field) were culled. Also, at the 
distance of HCG 22a and 90c, the number of globular clusters brighter 
than $V=22$ is negligible and hence all objects brighter than this limit 
were removed from the object catalogs. 

Besides the magnitude criteria, all objects that lie within the masked 
regions (such as near the galaxy centers and regions containing bright 
galaxies) were eliminated from the object catalogs. In order to improve 
the S/N of detected globular clusters over background galaxies, all objects 
detected more than $2.13\arcmin$ from the center of HCG 22a were discarded 
(at this radius the density of globular cluster candidates becomes 
negligible; see section \ref{sec:radial}). For HCG 90c, a cutoff radius 
of $2.58\arcmin$ was chosen.

Figures~\ref{22a_xy} and \ref{90c_xy} depict the spatial distribution of 
globular cluster candidates for HCG 22a and HCG 90c. A concentration of 
star-like objects surrounding each galaxy center is clearly seen.

Figures~\ref{22a_vr}, \ref{90c_vr}, and \ref{bkg_vr} give the color-magnitude 
distribution for the galaxy and control fields. It appears from these 
figures that the distribution of stellar-like objects in HCG 22a and HCG 90c 
are generally more  concentrated at a specific color than those objects 
found in the control field. Due to the small number of detected objects 
and the very short color baseline, we have not imposed a color selection 
criteria.      
\section{Results}\label{sec:results}

\subsection{Globular Cluster Luminosity Function}

At a limiting magnitude of $V \simeq 24.6$ we sample only the brightest 
fraction of globular clusters present around HCG 22a and 90c. Consequently, 
it is necessary to extrapolate counts to include the more numerous faint 
globulars not detected in our images.

Globular cluster luminosity functions are universally well described by a 
Gaussian function \citep{har91} of the form
\begin{equation}
N(V)=A~exp\left[\frac{-(m-m_{o})
^{2}}{2\sigma^{2}}\right],
\end{equation}
where $N(V)$ is the number of globular clusters, $A$ is the amplitude, 
$m$ is the apparent magnitude of individual globular clusters, $m_{o}$ 
is the magnitude of the turnover, and $\sigma$ is the dispersion.

The globular cluster luminosity function (GCLF) for each galaxy was 
constructed from the statistical subtraction of star-like objects measured 
from the control field. Unresolved objects in the galaxy field will consist 
of true globular clusters, foreground stars, and faint background galaxies 
which have stellar-like profiles and were miss-classified by FOCAS. In fact, 
most of the contamination will be due to faint unresolved galaxies since 
one expects less than one foreground star per square arcminute at 
$V > 21$ \citep[e.g.,][]{rat85}. Since object detection and classification 
were carried out identically for both the galaxy and control fields, the 
difference in the number of star-like objects between the galaxy fields and 
the control field (corrected for completeness and normalized with respect 
to spatial coverage) should be, statistically, globular cluster candidates.

Fitting a Gaussian function to the GCLF data involves simultaneously fitting 
three parameters; the turnover, dispersion, and the amplitude. However, since 
the GCLF data for HCG 22a and 90c are highly uncertain near the completeness 
limit, it was decided not to attempt to simultaneously solve for all three 
parameters. The absolute magnitude of the turnover of the GCLF is observed 
to be roughly the same for all galaxies, regardless of morphological type 
or local environment \citep{har91,jac92}. We have therefore adopted 
$M_{V}= -7.4 \pm 0.2$ for the absolute magnitude of the turnover, which is 
consistent with previous studies \citep[e.g.,][]{fle95}. Since the dispersion 
of the Gaussian function has been shown to have only a narrow range of values 
centered on $\sigma_{gauss}=1.4$ \citep{har91,jac92}, we have decided to 
adopt this value. A non-linear weighted least-squares fit of the GCLF of 
HCG 22a, using $ M_{V}= -7.4 \pm 0.2$, $\sigma_{gauss}=1.4$ and a distance 
of 54.1 Mpc, gives 
$A=38.4\pm 8.1$ with $\chi^{2}_{\nu}=1.4$. For HCG 90c, we found 
$A=51.3\pm 7.6$ and $\chi^{2}_{\nu}=1.9$, using a distance of 53.9 Mpc. 
Figure~\ref{22a_gclf} and 
\ref{90c_gclf} show the GCLF, with the best-fit Gaussian function, for 
HCG 22a and HCG 90c, respectively. Counts of stellar-like objects in the 
galaxy and control fields are given in Table~\ref{22agclf_data} and 
\ref{90cgclf_data} for HCG 22a and HCG 90c, respectively.

\subsection{Radial Distribution}\label{sec:radial}

In addition to correcting globular cluster counts for incomplete 
sampling of the GCLF as discussed above, it is also necessary to 
correct the counts for the limited areal coverage around each 
galaxy.   

The globular cluster radial distribution was measured by dividing the 
galaxy images into several annuli centered on each galaxy. For each annulus, 
the total number of globular cluster candidates were corrected for 
incompleteness and subtracted by the properly normalized control field counts.

The radial distribution was characterized using a power-law of the form, 
$\sigma =AR^{\alpha}$ \citep[e.g.,][]{fle95}. Using weighted non-linear 
least-squares, the best-fit values for the radial distribution of HCG 22a 
was determined to be $A=7.2\pm 1.28$ and $\alpha=-2.01\pm 0.30$, with 
$\chi^{2}_{\nu}=1.3$. Figure~\ref{radial_22a} depicts the radial distribution 
of globular cluster candidates for HCG 22a along with the best-fit power-law.

For HCG 90c, the best-fit values for the radial distribution was found to 
be $A=13.99\pm 1.42$ and $\alpha=-1.20\pm 0.16$, with $\chi^{2}_{\nu}=1.5$. 
Figure~\ref{radial_90c} depicts the radial profile of the globular cluster 
candidates, along with the best-fit power-law.

A useful comparison can be made between the spatial extent of the globular 
cluster system and the stellar halo of the parent galaxy. 
Figure~\ref{radSB_22a} shows the comparison of the radial profile of the 
globular cluster system and the surface brightness for HCG 22a. The surface 
brightness data has been taken from \citet{fra89} and has been scaled 
vertically to match the GC profile near its center. A fit to the radial 
dependence of the halo surface brightness, $\mu$, gives, $\mu\sim R^{-2.3}$, 
which is consistent with the slope of the radial profile of the GCs 
($\alpha =-2.01\pm0.3$). Thus, the globular cluster system is as spatially 
extended as the halo light of HCG 22a, which is consistent with other 
studies where the GC profiles are either similar to or more extended than 
the galaxy halo \citep[e.g.,][]{fle95}.  

The surface brightness profile of HCG 90c was determined from 
\citet{pen94} and is compared with the radial distribution of its globular 
cluster population in Figure~\ref{radSB_90c}. The surface brightness 
profile has been scaled vertically to match the globular cluster 
distribution. A fit to the halo profile gives a slope of $\mu\sim R^{-1.5}$ 
and hence, the globular cluster population is spatially more extended than 
the galaxy halo. We also note that the halo light of HCG 90c is not well 
defined due to the high surface brightness of the intra-group light 
\citep{whi01}. 

\subsection{Specific Frequency}

As mentioned in Section~\ref{sec:intro}, a useful way of quantifying 
the globular cluster population of different galaxies is to compare 
values of specific frequency, $S_{N}=N_{tot}10^{0.4(M_{V}+15)}$. The first 
step in calculating $S_{N}$ is to determine the total number of globular 
clusters, $N_{tot}$. The power-law radial distribution function, 
$\sigma =AR^{\alpha}$, must be integrated from an inner to an outer radius 
in order to account for the globular clusters that were missed due to 
spatial incompleteness. Also, the number of globular clusters must be 
corrected for incomplete coverage of the luminosity function. 

\subsubsection{Hickson Compact Group 22a}

In order to integrate the power-law radial distribution function, inner 
and outer radial limits must be chosen. Globular cluster distributions 
normally do not extend to radii inward of 1 kpc from the center of the 
parent galaxy \citep[e.g.,][]{har91,bri91}, the most likely cause being 
tidal shocking and dynamical friction which have destroyed globular 
clusters at distances close to the galaxy center \citep[e.g.,][]{wei93}. 
For an outer radius, globular cluster systems generally do not extend 
past 50 kpc from the center of a typical elliptical galaxy 
\citep[e.g.,][]{har81,han86,fle95}. Thus for this study, an inner radius 
of 1 kpc and an outer radius of 50 kpc was adopted for HCG 22a and 90c. 
These radial limits for HCG 22a correspond to angular distances of 
$0.06\arcmin$ and $3.1\arcmin$. A consistency check is provided by the fact 
that the radial counts for HCG 22a and 90c appear to reach background levels 
approximately 50 kpc from the galaxy center.

The integral of the power-law density profile gives 
$N_{tot}=\int\sigma_{cl}\,2\pi R~dR$, where $\sigma_{cl}$ represents 
the number density of globular clusters. For HCG 22a we have 
$N_{tot}=2\pi A\int_{0.06\arcmin}^{3.18\arcmin} R^{\alpha + 1}dR$. 
Solving this equation gives the total number of globular clusters, 
$N_{tot} = 179\pm 54$ {\it before} including the correction for faint 
globular clusters that are below our detection limit. If the outer 
radius was set to a smaller value, say 25 kpc, the total number of globular 
clusters would be $\sim 148$ and hence, $N_{tot}$ is not extremely sensitive 
to our chosen value of an outer radius for a steep radial slope of 
$\alpha \sim -2.0$.

After correcting for the radial incompleteness of detected globular 
clusters, the incomplete coverage of the luminosity function must be 
accounted for. Since the luminosity function was assumed to be a Gaussian, 
the area under the Gaussian function that was accounted for will give a 
measure of the percentage of observed globular clusters. Integrating the 
curve from $-\infty$ to the photometric limit of the data ($V=24.6$), 
indicates that $11\pm 3\%$ of the distribution was observed. Correcting 
$N_{tot}$ for the unobserved part of the luminosity function gives 
$N_{tot}^{cor}=1590\pm 854$.

Once the total number of globular clusters was determined, the specific 
frequency was calculated from the absolute V magnitude of the parent 
galaxy. Using an absolute magnitude of $M_{V}= -22.31\pm 0.07$ as given by 
NED\footnote{The NASA/IPAC Extragalactic Database (NED) is operated by the 
Jet Propulsion Laboratory, California Institute of Technology, under contract 
with the National Aeronautics and Space Administration.}, the specific 
frequency was determined to be $S_{N} = 1.9\pm 1.0$ for HCG 22a (using  
$H_{0}= 75~\mbox{km}~\mbox{s}^{-1}\mbox{Mpc}^{-1}$, we find 
$S_{N} = 1.7\pm 0.6$).

\subsubsection{Hickson Compact Group 90c}

The specific frequency for HCG 90c was calculated in the analogous way as 
for HCG 22a. The total number of globular clusters, for the chosen radial 
limits of 1 kpc to 50 kpc ($0.07\arcmin$ to $3.5\arcmin$), was determined 
to be $N_{tot}= 270\pm 28$. Changing the outer radius to 25 kpc yields 
$N_{tot}\sim 150$ and thus the total number of globular clusters is 
sensitive to the choice of an outer radius because of the shallow radial
profile with $\alpha \sim -1.2$. 

Correcting for the unobserved part of the luminosity function, we determined 
that $13\pm 3\%$ of the total number of globular clusters were observed 
and that $N_{tot}^{cor}= 2136\pm 718$. Using an absolute magnitude of 
$M_{V}=-21.99\pm 0.03$, we find that $S_{N} = 3.4\pm 1.1$ for 
HCG 90c (using $H_{0}= 75~\mbox{km}~\mbox{s}^{-1}\mbox{Mpc}^{-1}$, we find 
$S_{N} = 4.4\pm 1.3$).

\section{Discussion}\label{sec:dis}

In this paper we have investigated the globular cluster population of 
two prominent elliptical galaxies, each located near the center of a 
Hickson compact group. The globular cluster population has been 
characterized by measuring the number of globular clusters per unit galaxy 
luminosity ($S_{N}$), radial distributions, and the globular cluster 
luminosity function. Although our sample is limited to only two compact 
group galaxies, we can draw some tentative conclusions. 

The most important aspect of this study is the determination of the 
specific frequency of HCG 22a and HCG 90c. As mentioned in 
section~\ref{sec:intro}, specific frequency values for field ellipticals 
generally range from $S_{N}= 2-3$, in contrast to values measured for 
cD galaxies at the center of rich clusters ($S_{N}=10-20$). Although 
$S_{N}$ seems to follow a general trend that as the local density increases 
the number of globular clusters per unit galaxy luminosity also increases, 
there are exceptions \citep[e.g.,][]{har95,woo00}. The evidence seems to 
suggest that a high density environment, at least at the current epoch, 
may be a necessary but not a sufficient condition for the formation of 
high $S_{N}$ galaxies. The relatively low values of $S_{N}$ for HCG 22a 
and 90c ($S_{N}=1.9\pm 1.0$ and $S_{N}=3.4\pm 1.1$, respectively) suggests 
that these galaxies formed their globular clusters at a time when they were 
located in a ``field-like'' low density environment. 

The exact nature of Hickson compact groups has been controversial since 
some studies have suggested that the majority of compact groups are chance 
projections of large filamentary structures \citep{her95} or a superposition 
of galaxies in a much larger group or poor cluster \citep{wal89}. N-body 
simulations have indicated that, given the small crossing times and low 
velocity dispersions \citep{hic92}, group member galaxies would combine 
to form a single elliptical galaxy on the order of a few crossing times 
(e.g., Barnes 1985; Mamon 1987; but see Mamon 2000) and several examples 
of isolated ellipticals which may be the surviving remnants of galaxy 
groups have been discovered (Mulchaey \& Zabludoff 1999). This 
suggests that 
if compact groups are gravitationally bound, group members may be just 
beginning to come together at the present epoch and that during the time 
of globular cluster formation, these galaxies were situated in a lower 
density environment. This scenario is consistent with the specific frequency 
values measured for HCG 22a and 90c.

Evidence to support the hypothesis that these groups are bound structures 
include; the detection of $H_{I}$ and X-rays \citep{wil87,pil95,pon96}, 
indicating significant dynamical evolution and galaxy interaction, the 
decrease in the relative number of spirals with decreasing group crossing 
times \citep{hic92}, and the presence of tidal tails and debris in some 
compact groups \citep{hun96}.

As mentioned in section~\ref{sec:intro}, both HCG 22 and 90 are weak X-ray 
emitters. Radial velocity measurements imply that both HCG 22a and 90c 
are located near the bottom of their group potential well. Also, HCG 90b 
is clearly interacting with HCG 90d \citep{pla99}. HCG 90 has also been 
shown to contain a spatially extended, diffuse, light component \citep{whi01}. 
These facts suggests 
that both HCG 22 and 90 are gravitationally bound structures.

In recent years, it has been shown that the globular cluster systems 
of many large elliptical galaxies contain two or more chemically 
distinct sub-populations, usually inferred on the basis of color 
information \citep[e.g.,][]{ost93,gei96,cot98}. A number of different 
theories have been proposed to explain the origin of these different 
globular cluster populations; among these are models in which two or more 
bursts of globular cluster formation occur \citep[e.g.,][]{ash92,for00} 
or accretion of globular clusters from other galaxies 
\citep[e.g.,][]{cot98,cot00}. Given the rather unique environments 
represented by compact groups, and the roles that mergers may have 
played, it would be interesting to ascertain whether Hickson compact 
group galaxies also exhibit multiple globular cluster populations. 
Unfortunately, the V-R information provided by 
this investigation does not give us a sufficiently long color 
baseline to explore this possibility. 

\section{Conclusions}

We have studied the globular cluster systems around the early-type galaxies 
HCG 22a (NGC 1199) and 90c (NGC 7173) using V and R band data obtained from 
ESO's NTT. Globular clusters have been detected as a statistical excess of 
star-like objects surrounding each galaxy. Object detection and 
classification was performed using FOCAS on galaxy-subtracted images.

The specific frequency of each galaxy was measured assuming a turnover 
and dispersion of the globular cluster luminosity function. We found that 
for HCG 22a, $S_{N}= 1.9 \pm 1.0$, and for HCG 90c, $S_{N}= 3.4 \pm 1.1$. 
The total number of globular clusters for HCG 22a was determined to be 
$N_{tot} = 1590\pm 854$ and for HCG 90c we found $N_{tot} = 2136 \pm 718$. 
We also derived the globular cluster surface density profiles and found 
that for HCG 22a, the underlying starlight is as extended as the GC system. 
For HCG 90c, we found that the GC system is more extended than the galaxy halo.

From the results of this study, it is clear that further observations 
are needed to help understand the formation of globular clusters and 
the nature of Hickson compact groups. Observations which clearly reach 
the turnover in the globular cluster luminosity function would be of 
great value since they would provide a more robust determination of 
the GCLF, radial distribution, and specific frequency. Observations 
design to detect any bimodality in the color distribution would provide 
invaluable information on the past history of the globular cluster systems.

Additional observations of a larger sample of compact group galaxies 
would allow the investigation of potential correlations between galaxy 
luminosity and $S_{N}$ (since the local environment would be the same 
for a pair of galaxies in the same group) or other factors such as X-ray 
luminosity, mass-to-light ratio, and distance from the group dynamical center.

\acknowledgments
The authors would like to thank the referee for useful comments and 
suggestions.
This research has made use of the NASA/IPAC Extragalactic Database (NED) 
which is operated by the Jet Propulsion Laboratory, California Institute 
of Technology, under contract with the National Aeronautics and Space 
Administration. WAB is grateful for the support of the Reinhardt and Sumner 
fellowships. MJW acknowledges support from NSF grant AST 00-71149 and 
from NSERC of Canada.

\clearpage

\figcaption[]{V band image of Hickson compact group 22a (NGC 1199). 
\label{22a_image}}
\figcaption[]{V band image of Hickson compact group 90 with HCG 90c located 
near the center position. \label{90c_image}}
\figcaption[]{Positions of globular cluster candidates surrounding HCG 22a 
with $V\le 24.6$. The central 70 pixel masked-out region is shown along 
with the galaxy center at $X=885$ and $Y=647$ pixels. Only objects having 
radii $70 \le R \le 441$ pixels are shown. Note the concentration of 
star-like objects around the center of HCG 22a. \label{22a_xy}}
\figcaption[]{Positions of globular cluster candidates surrounding HCG 90c 
with $V\le 24.6$. The central 60 pixel masked-out region is shown along with 
the galaxy center at $X=887$ and $Y=652$ pixels. Only those objects having 
radii $60 \le R \le 535$ pixels are shown, excluding the area in the lower 
left quadrant which was masked out due to the presence of HCG 90b and d. 
Note the concentration of star-like objects around the center of 
HCG 90c.\label{90c_xy}}
\figcaption[]{Color-magnitude diagram of the globular cluster candidates 
surrounding HCG 22a. Note the enhanced concentration of stellar-like objects 
near $V-R=0.6$. \label{22a_vr}}
\figcaption[]{Color-magnitude diagram of the globular cluster candidates 
surrounding HCG 90c. Note the enhanced concentration of stellar-like objects 
near $V-R=0.5$. \label{90c_vr}}
\figcaption[]{Color-magnitude diagram of stellar-like objects in the control 
field. Note that objects appear to be scattered to a greater extent than 
those found for HCG 22a and 90c. \label{bkg_vr}}
\clearpage
\figcaption[]{Globular cluster luminosity function of HCG 22a. The best-fit 
Gaussian function is shown as the dashed line with a turnover of 
$m_{V}= 26.3\pm 0.2$ and a dispersion of $\sigma= 1.4$. \label{22a_gclf}} 
\figcaption[]{Globular cluster luminosity function of HCG 90c. The best-fit 
Gaussian function is shown as the dashed line with a turnover of 
$m_{V}= 26.0 \pm 0.2$ and a dispersion of $\sigma = 1.4$. \label{90c_gclf}} 
\figcaption[]{Radial distribution of globular clusters around HCG 22a. The 
best-fit power-law function, $\sigma \sim R^{-2.01\pm 0.30}$, is shown as 
the dashed line. The radius is given as the geometric mean of the individual 
annulii,$<R>=\sqrt{R_{in}R_{out}}$, and $\sigma$ is the number of star-like 
objects per $\mbox{arcminute}^{2}$. \label{radial_22a}}
\figcaption[]{Radial profile of globular clusters surrounding HCG 90c. 
The best-fit power-law, $\sigma \sim R^{-1.20\pm 0.16}$ is shown as a 
dashed line. The radius and $\sigma$ are as defined for HCG 22a. 
\label{radial_90c}}
\figcaption[]{The radial profile of the globular cluster system of HCG 22a 
(dashed line) compared to the R-band galaxy surface brightness from 
\citet{fra89} (solid line). The surface brightness profile has been scaled 
vertically in order to match the GC profile near the center. From the figure, 
it is clear that the globular cluster system is as spatially extended as the 
halo light of the parent galaxy. The slope of the power-law fit to the GC 
profile is $\alpha = -2.01\pm 0.3$ and for the halo light the slope is 
$\alpha \sim -2.3$. \label{radSB_22a}}
\figcaption[]{The radial profile of the globular cluster system of HCG 90c 
(solid line) compared to the R-band galaxy surface brightness 
from \citet{pen94} (dashed line). The surface brightness profile has been 
scaled vertically in order to match the GC profile near the center. From 
the figure, it is clear that the globular cluster system is more spatially 
extended than the halo light of the parent galaxy. The slope of the 
power-law fit to the GC profile is $\alpha = -1.20\pm 0.16$ and for the 
halo light the slope is $\alpha \sim -1.5$. \label{radSB_90c}}

\clearpage

\begin{deluxetable}{lll}
\tabletypesize{\scriptsize}
\tablecaption{General properties of HCG 22a and HCG 90c
 \label{gal_data}}
\tablewidth{0pt}
\tablehead{
\colhead{} & \colhead{HCG 22a\tablenotemark{1}} & \colhead{HCG 90c}}
   
\startdata
$\alpha$ (1950) & 03 01 18.2 & 21 59 08.8 \\
$\delta$ (1950) & -15 48 30 & -32 12 58 \\
{\it l} (deg)&199.22 & 14.98 \\
{\it b} (deg)&-57.31 & -53.08 \\ 
Hubble Type & E2 & E0 \\
$V_{o}$ ($\mbox{km}~\mbox{s}^{-1}$)& 2705 & 2696 \\
$B_{T}$(mag) & 12.24 & 12.73 \\
B-R (mag) & 1.62 & 2.23 \\
\enddata
\tablenotetext{1}{Table information from \citet{hh94}.}
\end{deluxetable}

\clearpage

\begin{deluxetable}{llll}
\tabletypesize{\scriptsize}
\tablecaption{Globular Cluster Counts: HCG 22a
 \label{22agclf_data}}
\tablewidth{0pt}
\tablehead{
\colhead{V} & \colhead{$N_{gal.}$}& \colhead{$N_{bkg}$} &\colhead{N}}
   
\startdata
22.1 & $3.0\pm 1.7$ & $0.6\pm 0.4$ & $2.4\pm 1.8$\\
22.3 & $3.0\pm 1.7$ & $2.2\pm 0.8$ & $0.8\pm 1.9$\\
22.5 & $0.0\pm 0.0$ & $0.8\pm 0.5$ & $-0.8\pm 0.5$\\
22.7 & $5.0\pm 2.2$ & $1.4\pm 0.6$ & $3.6\pm 2.3$\\ 
22.9 & $7.0\pm 2.6$ & $2.8\pm 0.9$ & $4.2\pm 2.8$\\
23.1 & $3.1\pm 1.8$ & $0.6\pm 0.4$ & $2.5\pm 1.8$\\
23.3 & $9.3\pm 3.1$ & $1.6\pm 0.3$ & $7.6\pm 3.1$\\
23.5 & $8.1\pm 2.9$ & $2.8\pm 0.9$ & $5.4\pm 3.0$\\
23.7 & $14.2\pm 4.2$ & $4.4\pm 1.1$ & $9.7\pm 4.3$\\
23.9 & $16.1\pm 4.4$ & $7.2\pm 1.4$ & $9.0\pm 4.6$\\
24.1 & $10.6\pm 3.4$ & $10.0\pm 1.6$ & $0.6\pm 3.8$\\
24.3 & $35.4\pm 6.6$ & $8.8\pm 1.7$ & $26.6\pm 6.8$\\
24.5 & $30.2\pm 7.0$ & $12.0\pm 2.0$ & $18.2\pm 7.3$\\
\enddata
\tablecomments{Column 1 gives the mid-bin V magnitude. Columns 2 and 3 gives 
the raw counts of stellar-like objects in the galaxy and control field. 
Column 4 gives the background-corrected counts of stellar-like objects 
in the galaxy field.}
\end{deluxetable}

\clearpage

\begin{deluxetable}{llll}
\tabletypesize{\scriptsize}
\tablecaption{Globular Cluster Counts: HCG 90c
 \label{90cgclf_data}}
\tablewidth{0pt}
\tablehead{
\colhead{V} & \colhead{$N_{gal.}$}& \colhead{$N_{bkg}$} & \colhead{N}}
   
\startdata
22.1 & $6.0\pm 2.4$ & $0.6\pm 0.4$ & $5.4\pm 2.5$\\
22.3 & $6.0\pm 2.4$ & $2.5\pm 0.9$ & $3.5\pm 2.6$\\
22.5 & $8.0\pm 2.8$ & $0.9\pm 0.5$ & $7.1\pm 2.9$\\
22.7 & $6.0\pm 2.4$ & $1.6\pm 0.4$ & $4.4\pm 2.5$\\ 
22.9 & $13.0\pm 3.6$ & $3.1\pm 1.0$ & $9.9\pm 3.7$\\
23.1 & $15.0\pm 3.9$ & $0.6\pm 0.4$ & $14.4\pm 3.9$\\
23.3 & $6.0\pm 2.4$ & $1.9\pm 0.8$ & $4.1\pm 2.6$\\
23.5 & $10.0\pm 3.2$ & $3.1\pm 1.0$ & $6.9\pm 3.3$\\
23.7 & $20.0\pm 4.5$ & $5.0\pm 1.2$ & $15.0\pm 4.6$\\
23.9 & $22.0\pm 4.7$ & $8.1\pm 1.6$ & $13.9\pm 5.0$\\
24.1 & $26.0\pm 5.1$ & $11.2\pm 1.9$ & $14.8\pm 5.4$\\
24.3 & $50.1\pm 7.4$ & $9.9\pm 1.9$ & $40.2\pm 7.7$\\
24.5 & $38.8\pm 6.9$ & $13.5\pm 2.3$ & $25.4\pm 7.2$\\
\enddata
\tablecomments{Column 1 gives the mid-bin V magnitude. Columns 2 and 3 gives 
the raw counts of stellar-like objects in the galaxy and control field. 
Column 4 gives the background-corrected counts of stellar-like objects 
in the galaxy field.}
\end{deluxetable}

\clearpage
\begin{figure}
\plotone{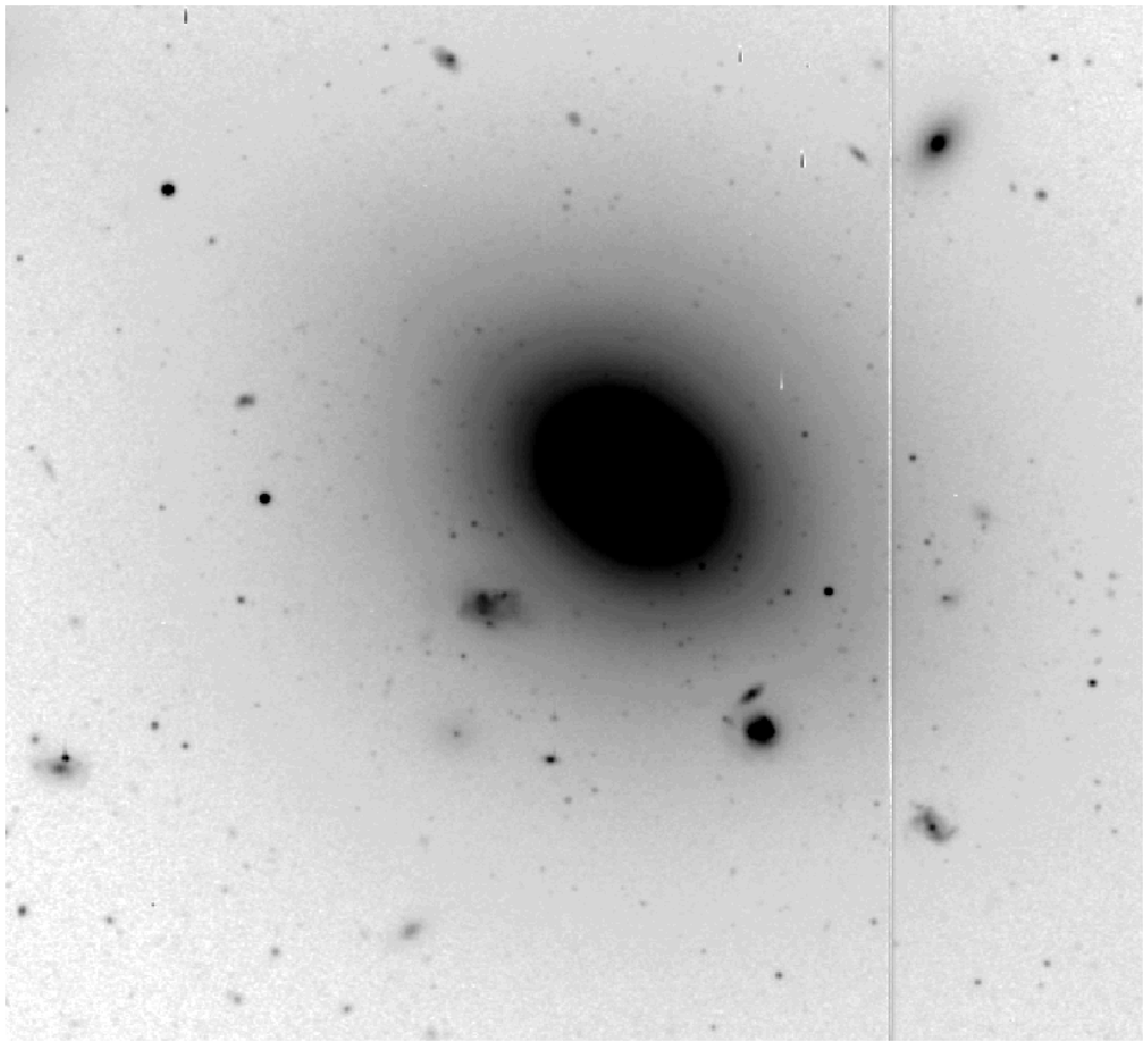}
\end{figure}
\clearpage
\begin{figure}
\plotone{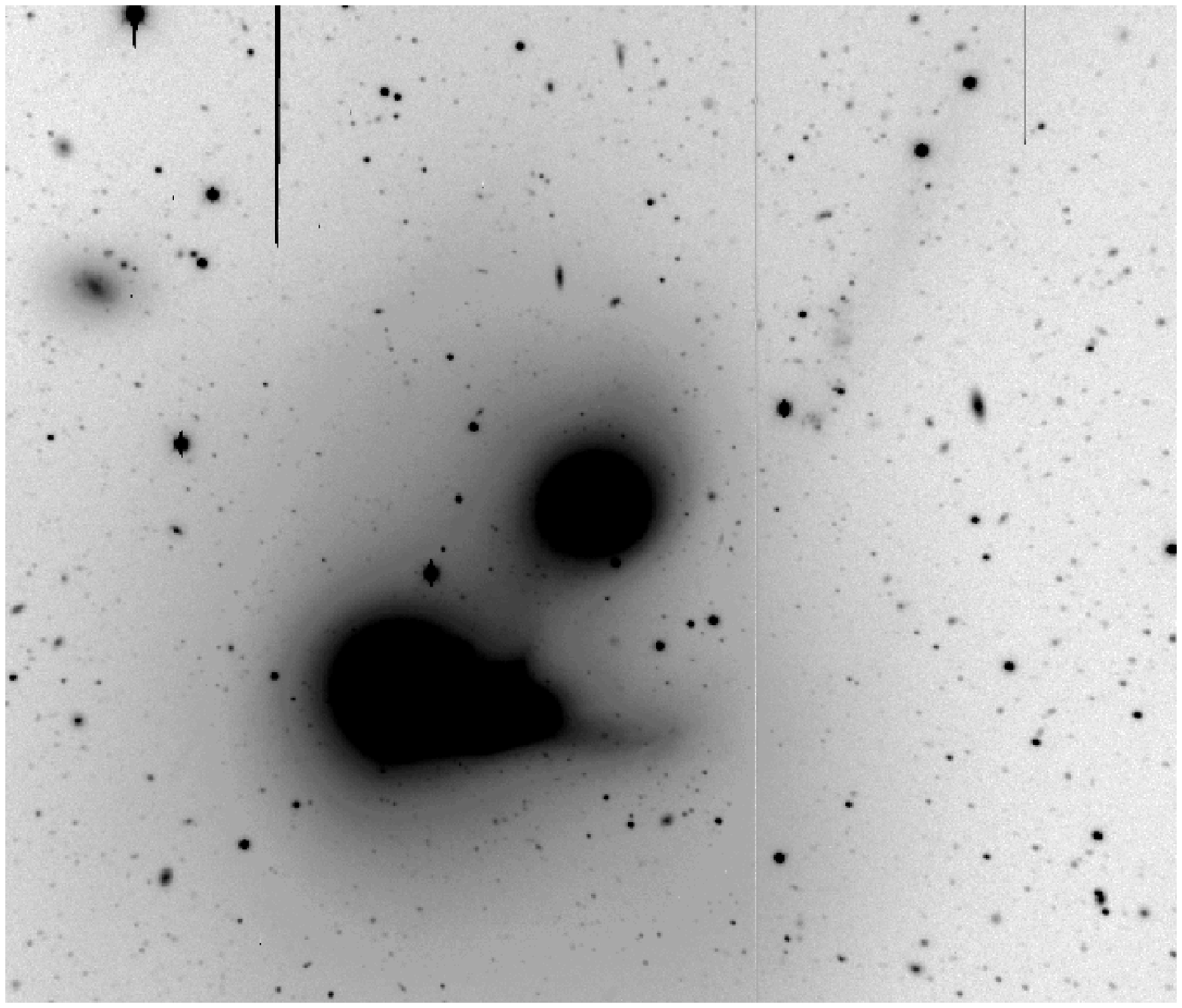}
\end{figure}
\clearpage
\begin{figure}
\plotone{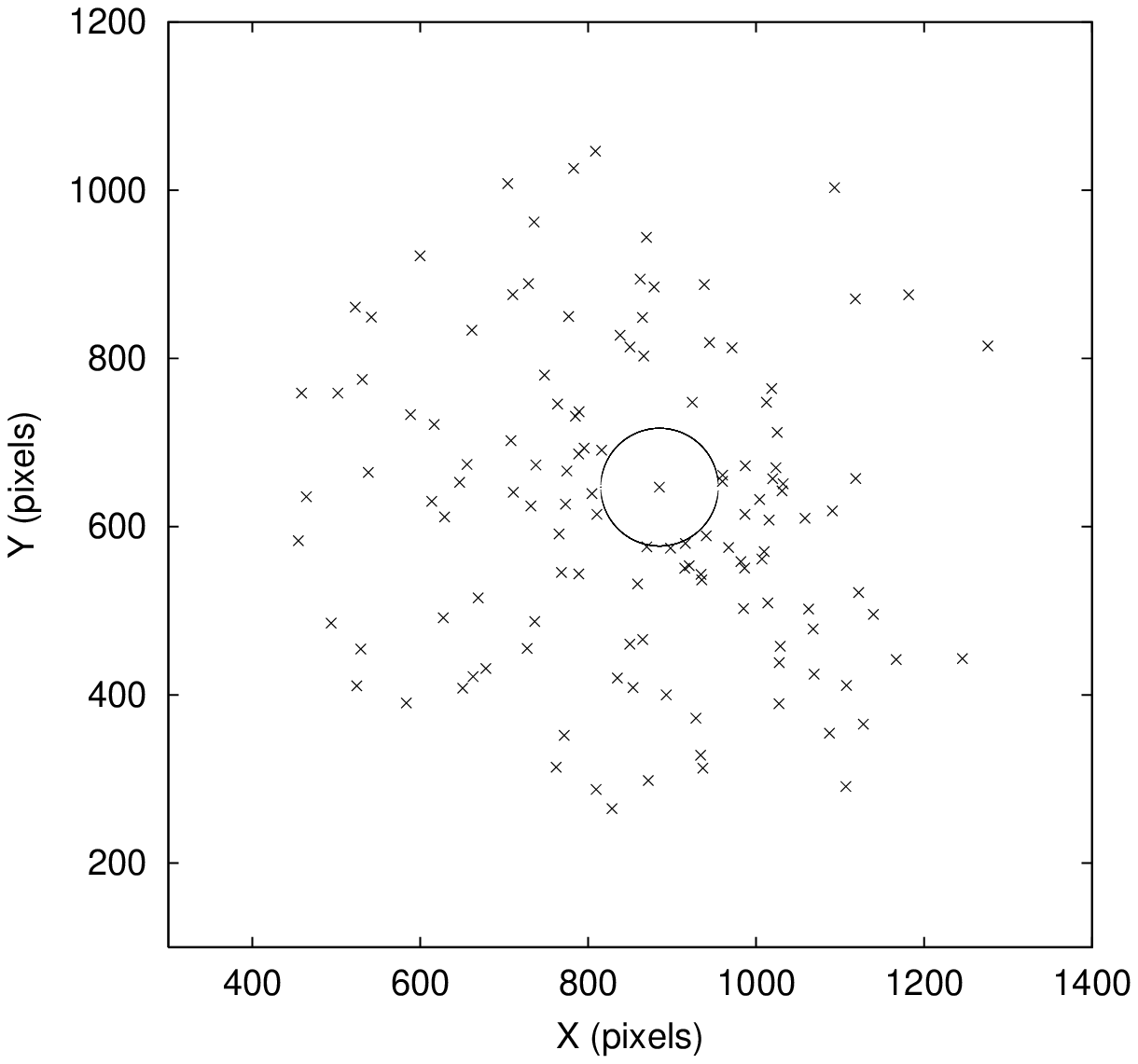}
\end{figure}
\clearpage
\begin{figure}
\plotone{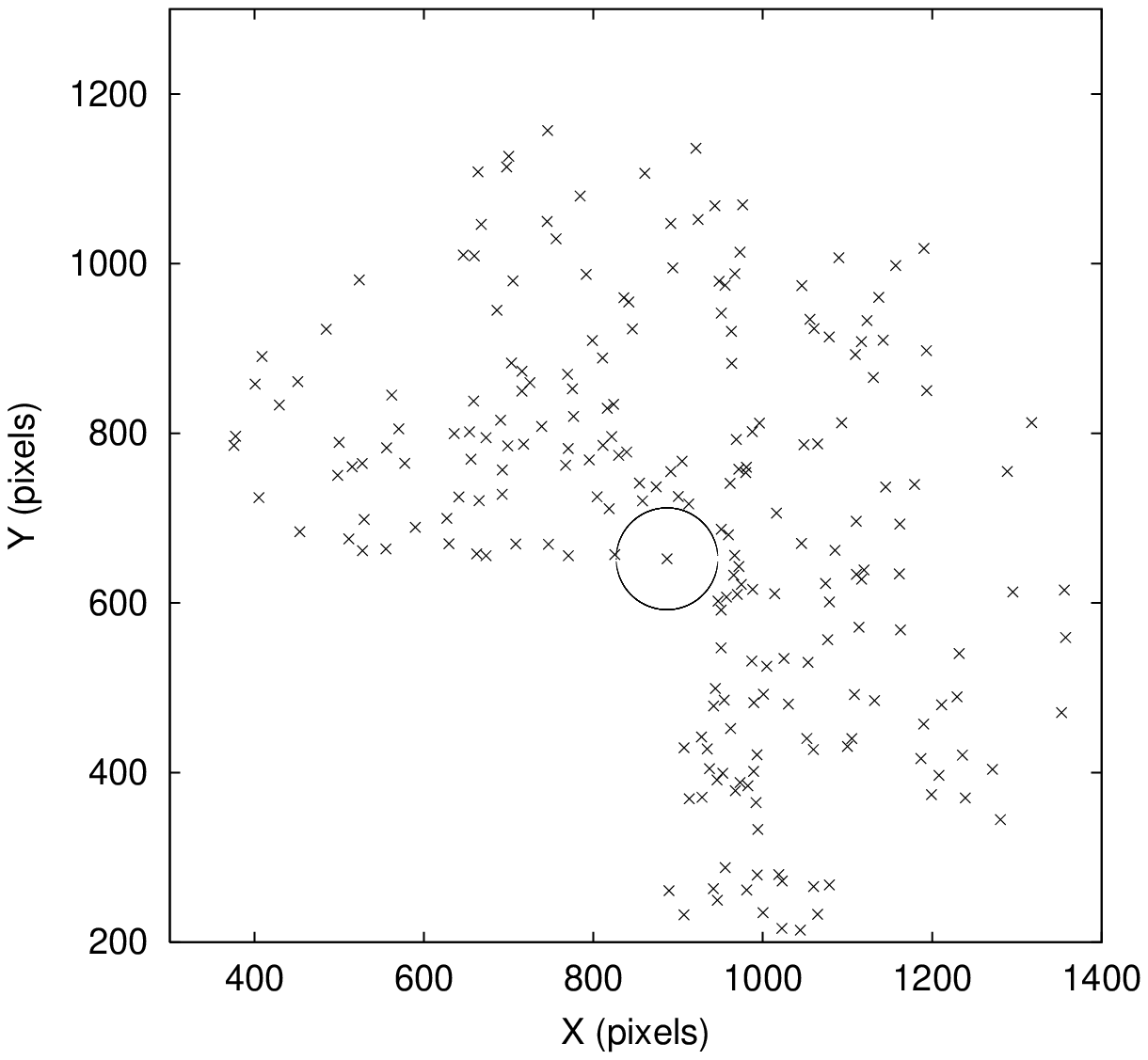}
\end{figure}
\clearpage
\begin{figure}
\plotone{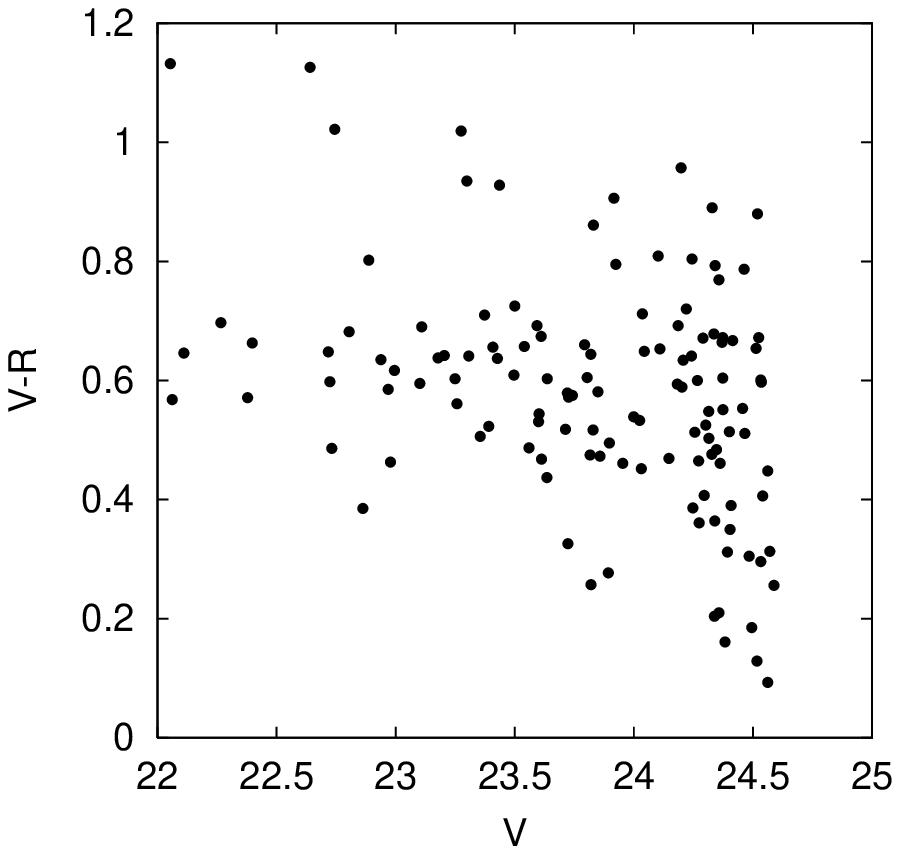}
\end{figure}
\clearpage
\begin{figure}
\plotone{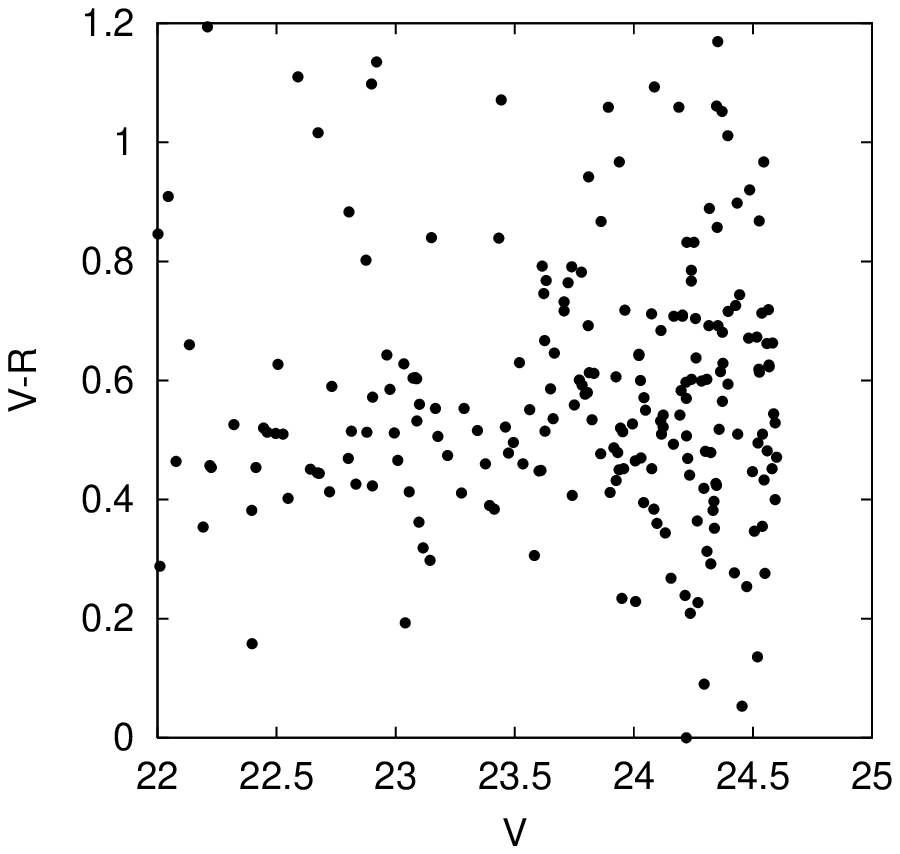}
\end{figure}
\clearpage
\begin{figure}
\plotone{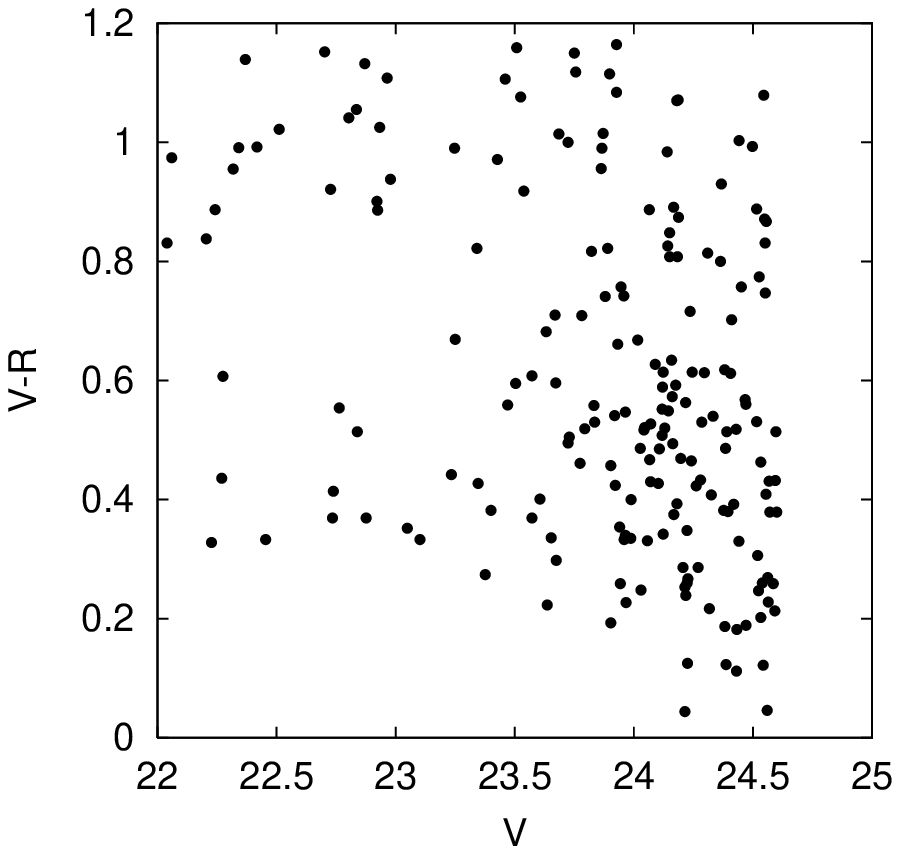}
\end{figure}
\clearpage
\begin{figure}
\plotone{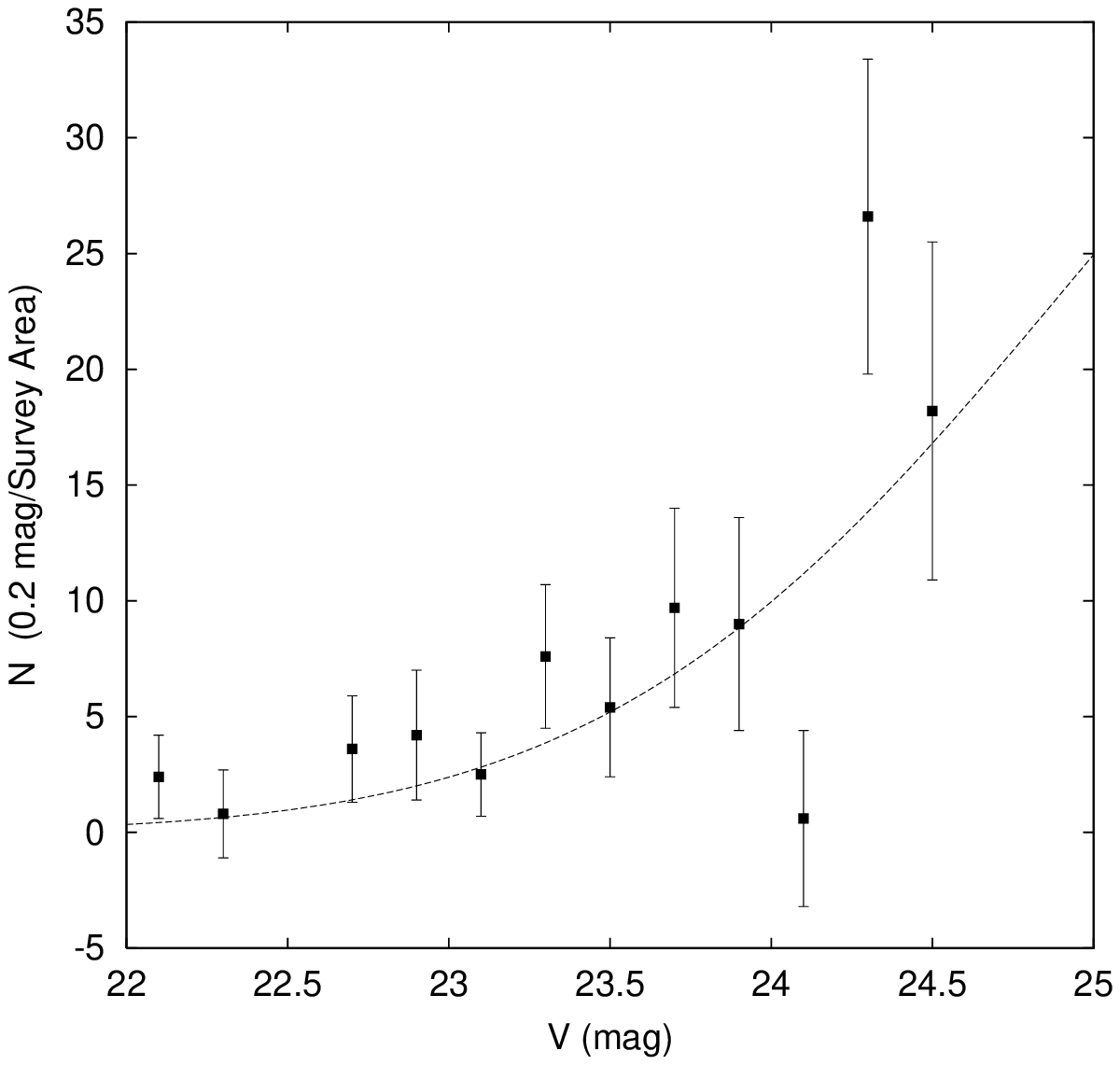}
\end{figure}
\clearpage
\begin{figure}
\plotone{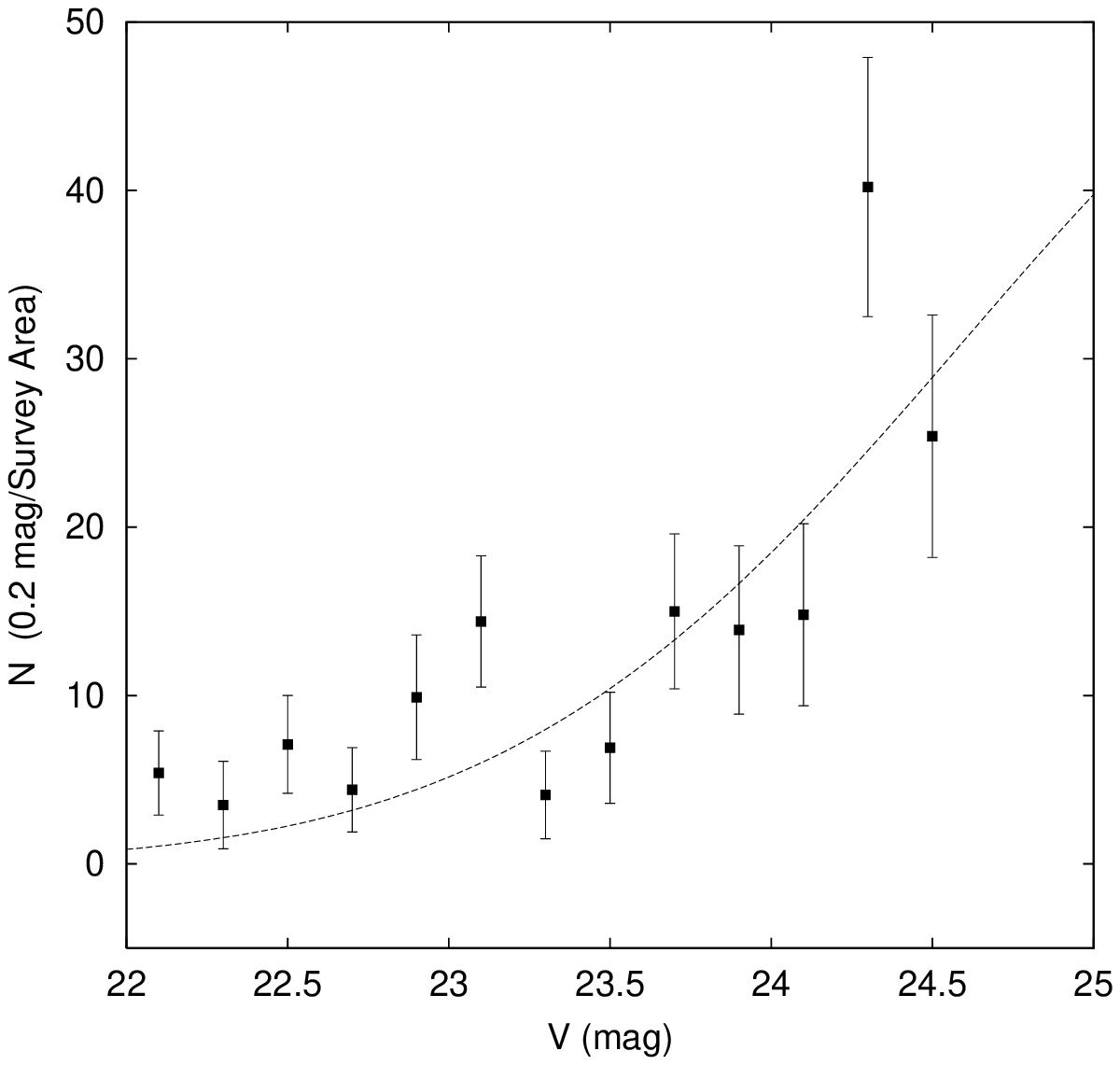}
\end{figure}
\clearpage
\begin{figure}
\plotone{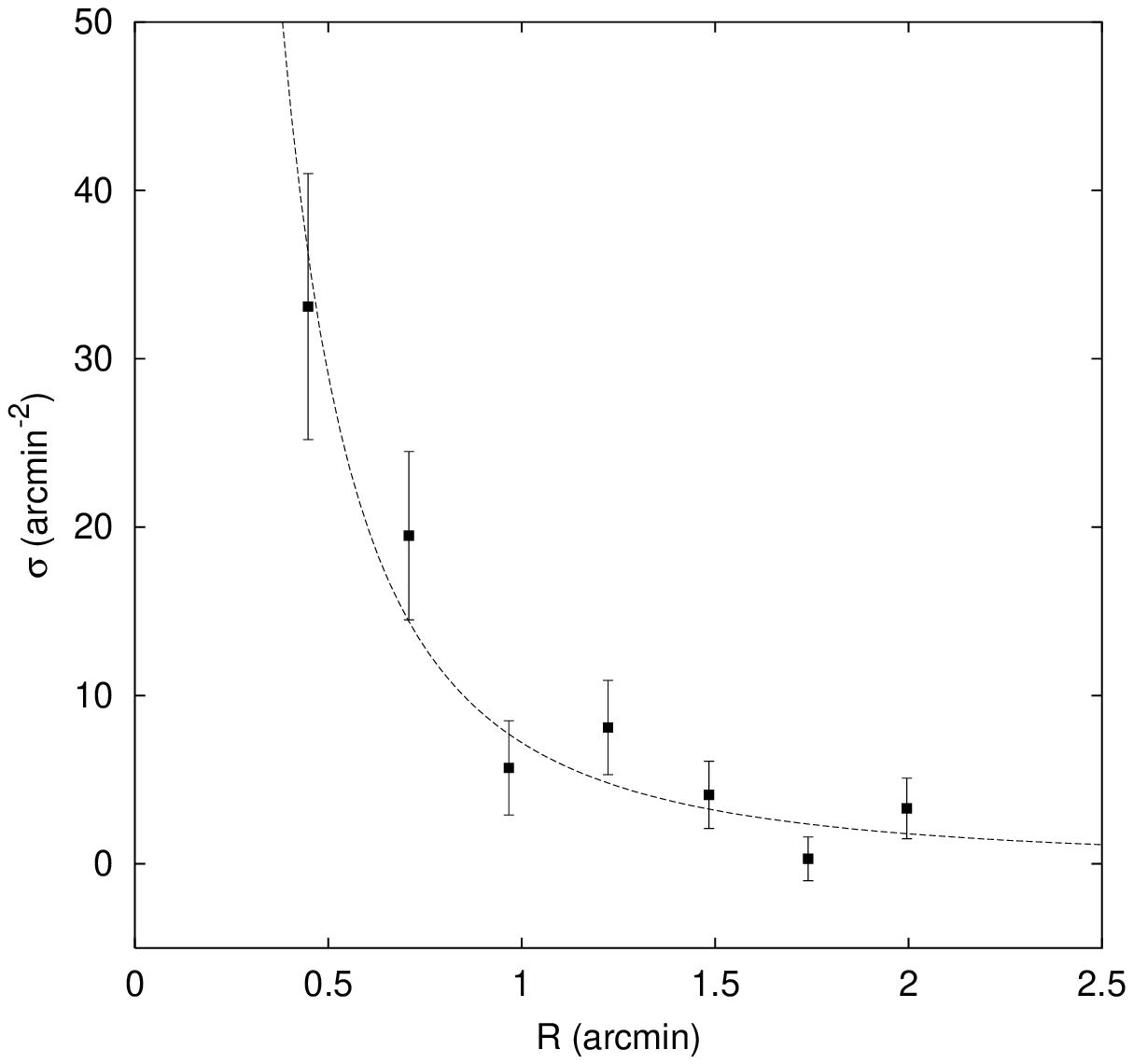}
\end{figure}
\clearpage
\begin{figure}
\plotone{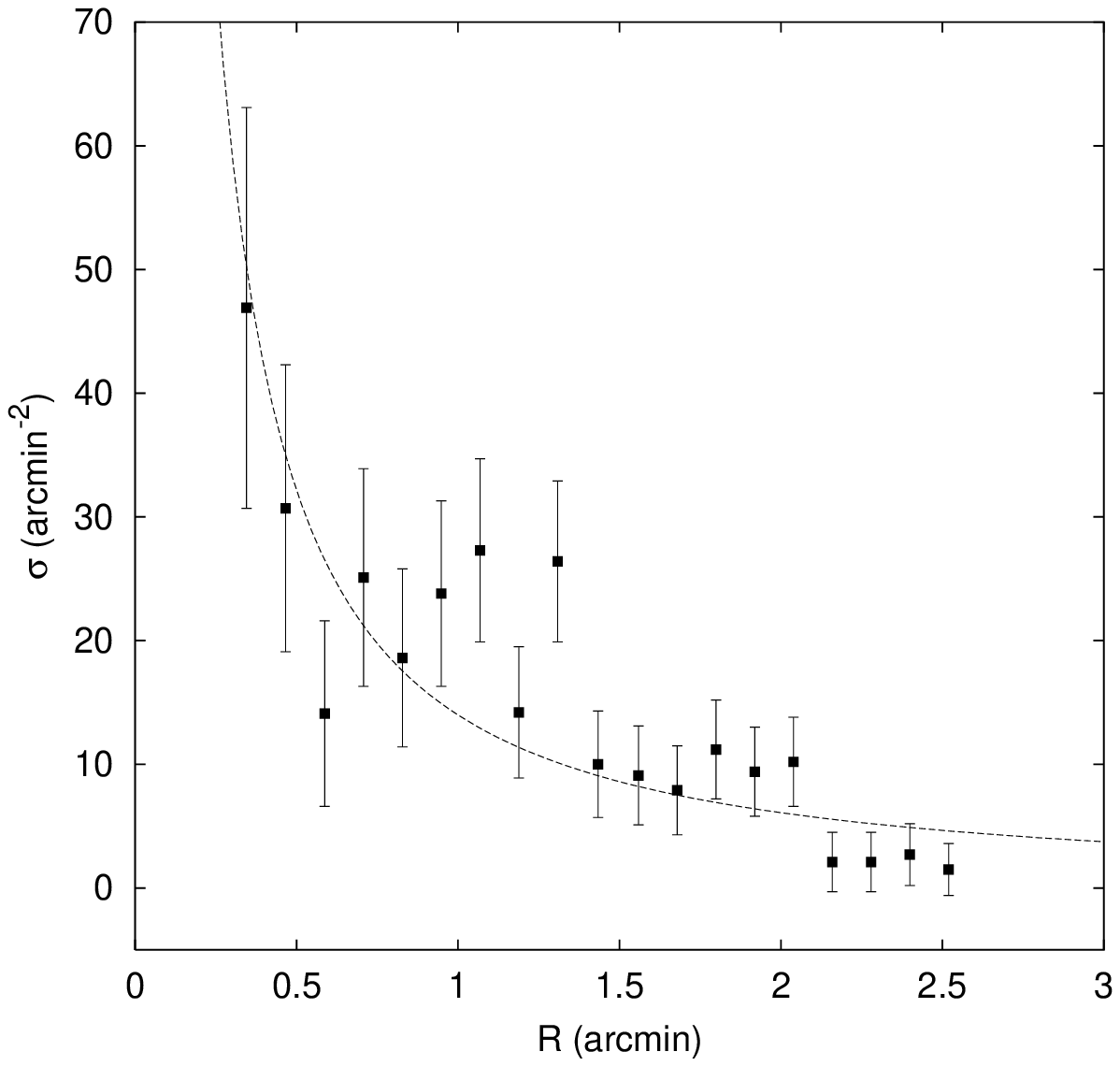}
\end{figure}
\clearpage
\begin{figure}
\plotone{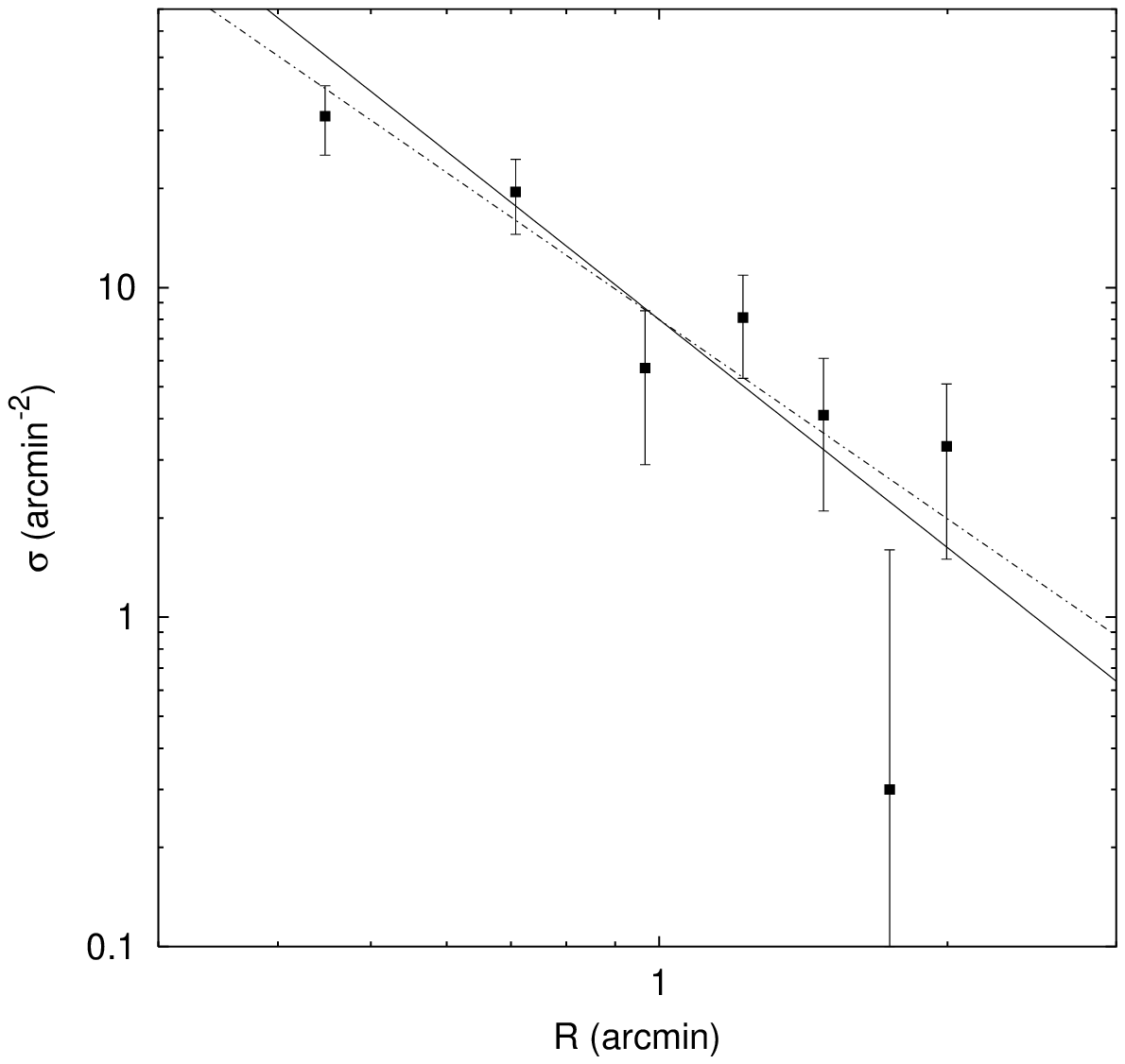}
\end{figure}
\clearpage
\begin{figure}
\plotone{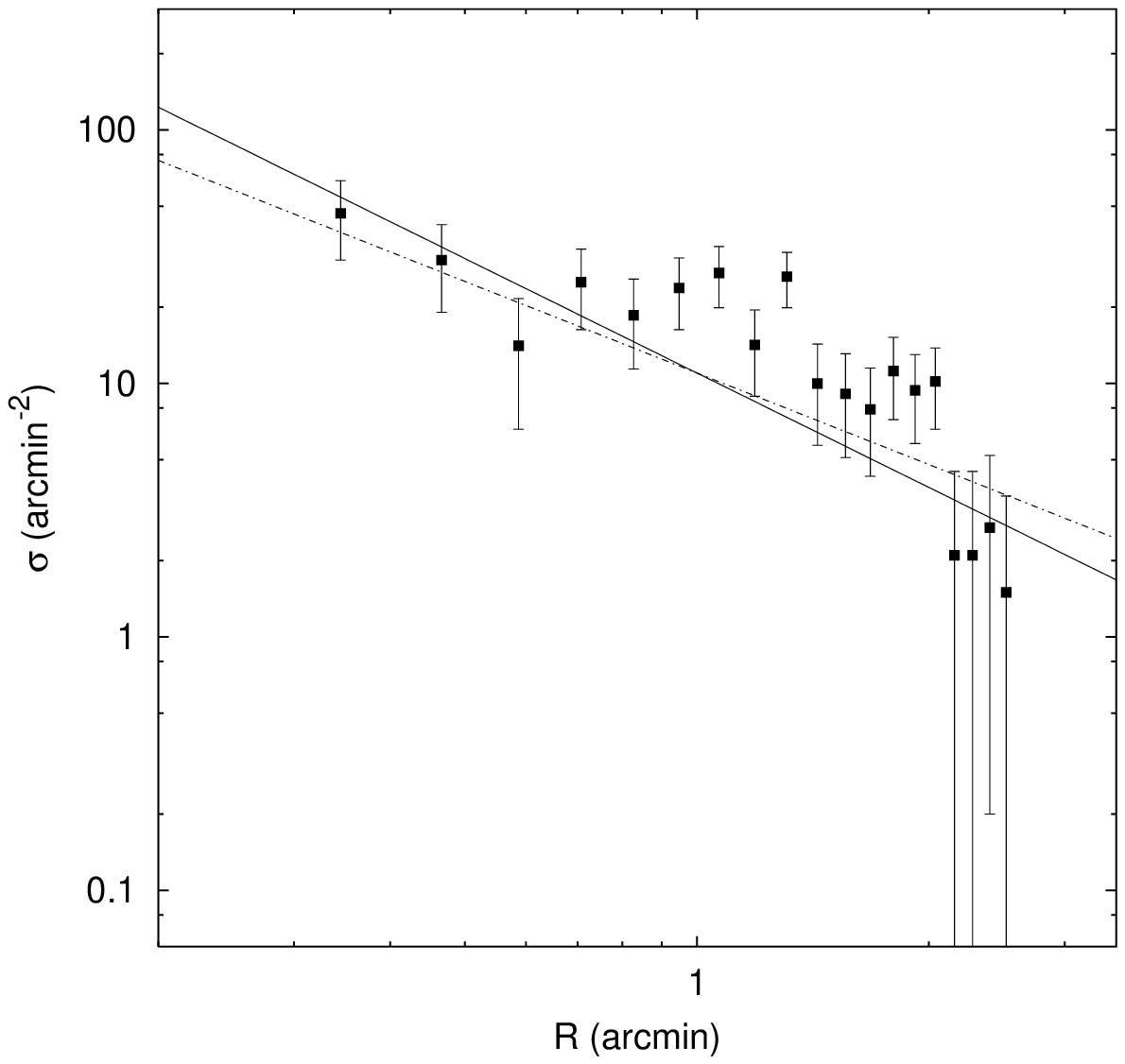}
\end{figure}

\end{document}